\documentstyle[epsfig]{mn}

\newif\ifAMStwofonts

\newcommand{\HI}   {H\,{\sc i}}

\newcommand{\HeII} {He\,{\sc ii}}

\newcommand{\CIII} {C\,{\sc iii}}
\newcommand{\CIV}  {C\,{\sc iv}}
\newcommand{\NIII} {N\,{\sc iii}}
\newcommand{\NaI}  {Na\,{\sc i}}
\newcommand{\CaII} {Ca\,{\sc ii}}
\newcommand{\Halpha}{H$\alpha$}
\newcommand{\Hbeta} {H$\beta$}

\ifoldfss
  \ifCUPmtlplainloaded \else
    \NewTextAlphabet{textbfit} {cmbxti10} {}
    \NewTextAlphabet{textbfss} {cmssbx10} {}
    \NewMathAlphabet{mathbfit} {cmbxti10} {} 
    \NewMathAlphabet{mathbfss} {cmssbx10} {} 
  \fi
  \ifAMStwofonts
    \ifCUPmtlplainloaded \else
      \NewSymbolFont{upmath} {eurm10}
      \NewSymbolFont{AMSa} {msam10}
      \NewMathSymbol{\upi}     {0}{upmath}{19}
      \NewMathSymbol{\umu}     {0}{upmath}{16}
      \NewMathSymbol{\upartial}{0}{upmath}{40}
      \NewMathSymbol{\leqslant}{3}{AMSa}{36}
      \NewMathSymbol{\geqslant}{3}{AMSa}{3E}

    \fi
  \fi
\fi 

\ifnfssone
  \newmathalphabet{\mathit}
  \addtoversion{normal}{\mathit}{cmr}{m}{it}
  \addtoversion{bold}{\mathit}{cmr}{bx}{it}

    \selectfont #1}}
  \newmathalphabet{\mathbfit} 
  \addtoversion{normal}{\mathbfit}{cmr}{bx}{it}
  \addtoversion{bold}{\mathbfit}{cmr}{bx}{it}
  \newmathalphabet{\mathbfss} 
  \addtoversion{normal}{\mathbfss}{cmss}{bx}{n}
  \addtoversion{bold}{\mathbfss}{cmss}{bx}{n}
  \ifAMStwofonts
    \ifCUPmtlplainloaded \else
      \UseAMStwoboldmath
      \makeatletter
      \new@mathgroup\upmath@group
      \define@mathgroup\mv@normal\upmath@group{eur}{m}{n}
      \define@mathgroup\mv@bold\upmath@group{eur}{b}{n}
      \edef\UPM{\hexnumber\upmath@group}
      \new@mathgroup\amsa@group
      \define@mathgroup\mv@normal\amsa@group{msa}{m}{n}
      \define@mathgroup\mv@bold\amsa@group{msa}{m}{n}
      \edef\AMSa{\hexnumber\amsa@group}
      \makeatother
      \mathchardef\upi="0\UPM19
      \mathchardef\umu="0\UPM16
      \mathchardef\upartial="0\UPM40
      \mathchardef\leqslant="3\AMSa36
      \mathchardef\geqslant="3\AMSa3E
    \fi
  \fi
\fi 

\ifnfsstwo
  \DeclareMathAlphabet{\mathbfit}{OT1}{cmr}{bx}{it}
  \SetMathAlphabet\mathbfit{bold}{OT1}{cmr}{bx}{it}
  \DeclareMathAlphabet{\mathbfss}{OT1}{cmss}{bx}{n}
  \SetMathAlphabet\mathbfss{bold}{OT1}{cmss}{bx}{n}
  \ifAMStwofonts
    \ifCUPmtlplainloaded \else
      \DeclareSymbolFont{UPM}{U}{eur}{m}{n}
      \SetSymbolFont{UPM}{bold}{U}{eur}{b}{n}
      \DeclareSymbolFont{AMSa}{U}{msa}{m}{n}
      \DeclareMathSymbol{\upi}{0}{UPM}{"19}
      \DeclareMathSymbol{\upartial}{0}{UPM}{"40}
      \DeclareMathSymbol{\leqslant}{3}{AMSa}{"36}
      \DeclareMathSymbol{\geqslant}{3}{AMSa}{"3E}
    \fi
  \fi
\fi 

\ifCUPmtlplainloaded \else
  \ifAMStwofonts \else 
    \def\upi{\pi}
    \def\umu{\mu}
    \def\upartial{\partial}
  \fi
\fi

\title[Optical studies of XTE J2123--058 -- II.  Spectroscopy] 
{Optical studies of the X-ray transient XTE J2123--058 -- II.  
Phase-resolved spectroscopy}
\author[R. I. Hynes et al.]
       {R. I. Hynes$^{1,3}$\thanks{e-mail: rih@astro.soton.ac.uk}, P. A. Charles$^{2,3}$, C. A. Haswell$^1$,
       J. Casares$^4$, C. Zurita$^4$, 
       \newauthor M. Serra-Ricart$^4$\\
$^1$Department of Physics and Astronomy, The Open University, Walton
    Hall, Milton Keynes, MK7 6AA\\
$^2$Astrophysics, Nuclear and Astrophysics Laboratory,
    Keble Road, Oxford, OX1 3RH\\
$^3$Department of Physics and Astronomy, University of Southampton, 
    Southampton, SO17 1BJ\\
$^4$Instituto de Astrof\'\i{}sica de Canarias, 38200 La Laguna,
    Tenerife, Spain} 
\date{Accepted ?.
      Received ?;
      in original form ?}

\pagerange{\pageref{firstpage}--\pageref{lastpage}}
\pubyear{2000}

\begin{document}
\maketitle
%
%
\begin{abstract}
We present time-resolved spectroscopy of the soft X-ray transient
XTE\,J2123--058 in outburst. A useful spectral coverage of
3700--6700\,\AA\ was achieved spanning two orbits of the binary, with
single epoch coverage extending to $\sim9000$\,\AA.  The optical
spectrum approximates a steep blue power-law, consistent with emission
on the Rayleigh-Jeans tail of a hot black body spectrum.  The
strongest spectral lines are \HeII\ 4686\,\AA\ and \CIII/\NIII\
4640\,\AA\ (Bowen blend) in emission.  Their relative strengths
suggest that XTE\,J2123--058 was formed in the Galactic plane, not in
the halo.  Other weak emission lines of \HeII\ and \CIV\ are present
and Balmer lines show a complex structure, blended with \HeII.  \HeII\
4686\,\AA\ profiles show a complex multiple S-wave structure with the
strongest component appearing at low velocities in the lower-left
quadrant of a Doppler tomogram.  \Halpha\ shows transient absorption
between phases 0.35--0.55.  Both of these effects appear to be
analogous to similar behaviour in SW~Sex type cataclysmic variables.
We therefore consider whether the spectral line behaviour of
XTE\,J2123--058 can be explained by the same models invoked for those
systems.
\end{abstract}
%
%
\begin{keywords}
accretion, accretion discs -- binaries: close -- stars: individual: 
XTE J2123--058
\end{keywords}
%
%
\section{Introduction}
\label{IntroSection}
The soft X-ray transient (SXT) XTE\,J2123--058 was discovered by the
All Sky Monitor (ASM) on the {\it RXTE} satellite, and confirmed by
the Proportional Counter Array (PCA), on 1998 June 27 (Levine, Swank
\& Smith 1998).  The object was promptly identified with a 17th
magnitude blue star with an optical spectrum typical of SXTs in
outburst (Tomsick et al.\ 1998a).  The discovery of apparent type-I
X-ray bursts (Takeshima \& Strohmayer 1998) confirmed that the
compact object was a neutron star and detection of twin high frequency
quasiperiodic oscillations (Homan et al.\ 1998, 1999; Tomsick et al.\
1999), as seen in other neutron star LMXBs (van der Klis 1999 and
references therein) is consistent with this.  Casares et al.\ (1998)
reported the presence of a strong optical modulation and attributed
this to an eclipse.  The orbital period was subsequently determined to
be 6.0\,hr both photometrically (Tomsick et al.\ 1998b; Ilovaisky \&
Chevalier 1998) and spectroscopically (Hynes et al.\ 1998).  Tomsick
et al.\ (1998b) suggested that the 0.9\,mag modulation is due to the
changing aspect of the heated companion in a high inclination system,
although partial eclipses appear also to be superimposed on this
(Zurita, Casares \& Hynes 1998).  Ilovaisky \& Chevalier (1998)
reported that the light curve shows a further 0.3\,mag modulation on a
7.2\,d period which they suggested was associated with disc precession.
By 1998 August 26 XTE\,J2123--058 appeared to have reached optical
quiescence, at a magnitude of $R\sim 21.7$ (Zurita \& Casares 1998;
Zurita et al.\ 2000, hereafter Paper I).

XTE\,J2123--058 is an important object to study for several reasons.
Firstly, neutron star transients with low-mass companions (in contrast
to the high-mass Be neutron star X-ray transients and the low-mass
black hole X-ray transients) are relatively rare; most neutron star
low-mass X-ray binaries (LMXBs) are persistently active.  It is
important to compare neutron star transients with the apparently more
common black hole systems in order to determine how the presence or
absence of a compact object hard surface affects the behaviour of the
system.  Such a comparison can also be used to test models for the
formation and evolution of SXTs which predict different evolutionary
histories for black hole and neutron star transients (King, Kolb \&
Szuszkiewicz 1997 and references therein).  Secondly, XTE\,J2123--058
has a high orbital inclination, and possibly exhibits partial
eclipses.  High inclination systems provide excellent tools with which
to probe the system geometry.  Thirdly, XTE\,J2123--058 is at high
Galactic latitude ($-36\fdg2$) and distant ($d=8\pm3$\,kpc, Paper I),
placing it in the Galactic halo; Tomsick et al.\ (1999) estimate a
distance of at least 2.6\,kpc below the Galactic plane.  For
comparison, the mean altitude of the low-mass X-ray binary
distribution (i.e.\ $z_{\rm rms}$) is only 1\,kpc (van Paradijs \&
White 1995).  We must therefore ask if this system is intrinsically a
halo object, or if it was formed in the disc and then kicked out at
high velocity when the neutron star was formed.  This issue is also
relevant to LMXB evolution (King \& Kolb 1997).

In this paper we present the results of a spectrophotometric study of
XTE\,J2123--058 using the William Herschel Telescope (WHT), La Palma.
We will address the questions and possibilities raised above.
Our photometric observations were described Paper I.
%
%
\section{Data reduction}
\label{ReductionSection}
We observed XTE\,J2123--058 using the ISIS dual-beam spectrograph on
the 4.2-m William Herschel Telescope (WHT) on the nights of 1998 July
18--20.  A log of exposures is given in Table \ref{WHTTable}.  The
R300B grating combined with an EEV $4096 \times 2048$ CCD gave an
unvignetted wavelength range of $\sim$4000--6500\,\AA\ with some
useful data outside this range.  An 0.7\,arcsec slit was used on July
19/20 and 1.0\,arcsec on July 20/21.  On both of these nights the slit
was aligned to pass through a second comparison star.  An additional
pair of exposures were taken on July 18/19, both using an 0.7\,arcsec
slit, with R300B/EEV10 on the blue arm and R158R/TEK2 on the red arm.
No flux calibration was attempted for these latter spectra and flat
fields were not available for the blue side spectrum.

\begin{table*}
\caption{Log of WHT spectroscopic observations of XTE\,J2123--058.
Each line represents a group of consecutive observations during which
autoguiding maintained the {\em red} images of the stars at fixed positions
on the slit.}
\label{WHTTable}
\begin{center}
\begin{tabular}{lccccc}
\noalign{\smallskip}
\hline
\noalign{\smallskip}
Date      & UT Range & Exposures & Number of & Wavelength  & Resolution \\
1998 July &          & Time (s)  & Exposures & range (\AA) & (\AA)      \\
\noalign{\smallskip}
18/19 & 02:40--03:00 & 1200 & 1 & 3750--6250 & 2.9 \\
      & 02:40--03:00 & 1200 & 1 & 6153--9124 & 3.4 \\
19/20 & 23:38--05:45 & 1200 &16 & 3750--6800 & 2.9 \\
20/21 & 23:46--01:29 & 1200 & 5 & 3750--6800 & 4.1 \\
      & 02:41--04:48 & 1200 & 6 & 3750--6800 & 4.1 \\
      & 05:10--05:30 & 1200 & 1 & 3750--6800 & 4.1 \\
      & 05:35--05:37 & 100  & 1 & 3750--6800 & 5.6 \\
\noalign{\smallskip}  
\hline
\end{tabular}
\end{center}
\end{table*}

\begin{figure}
\epsfig{angle=90,width=3in,file=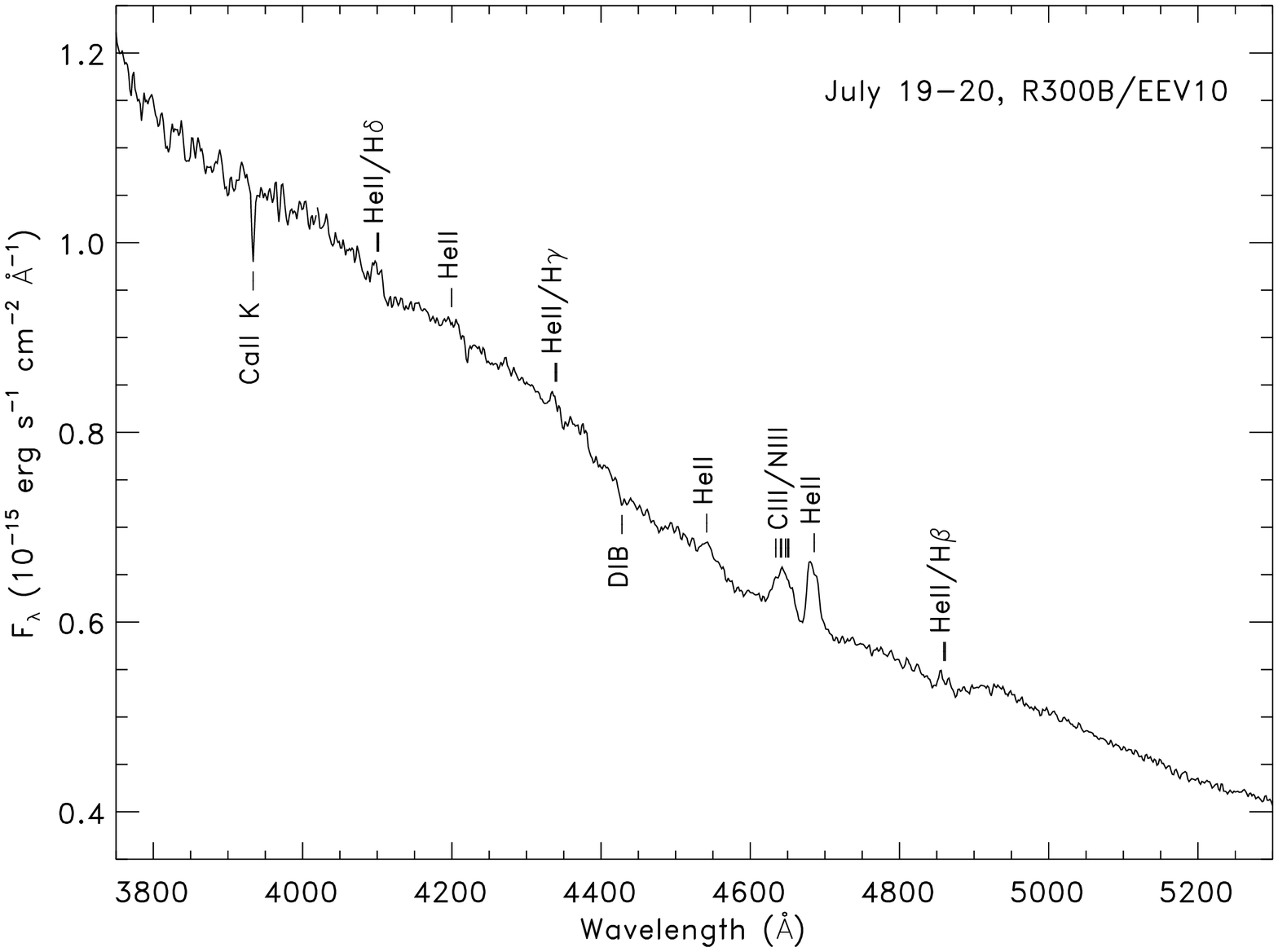}
\epsfig{angle=90,width=3in,file=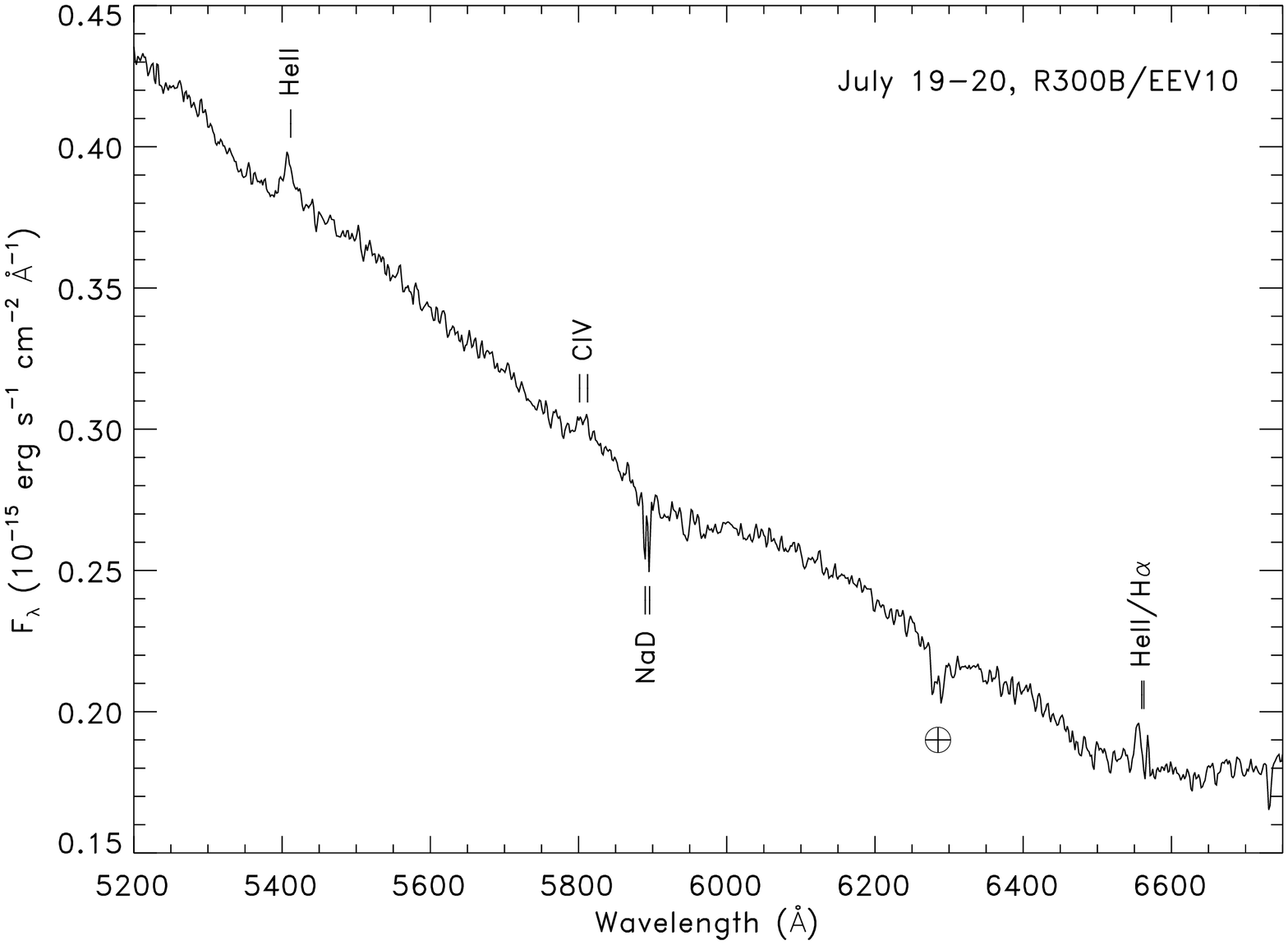}
\epsfig{angle=90,width=3in,file=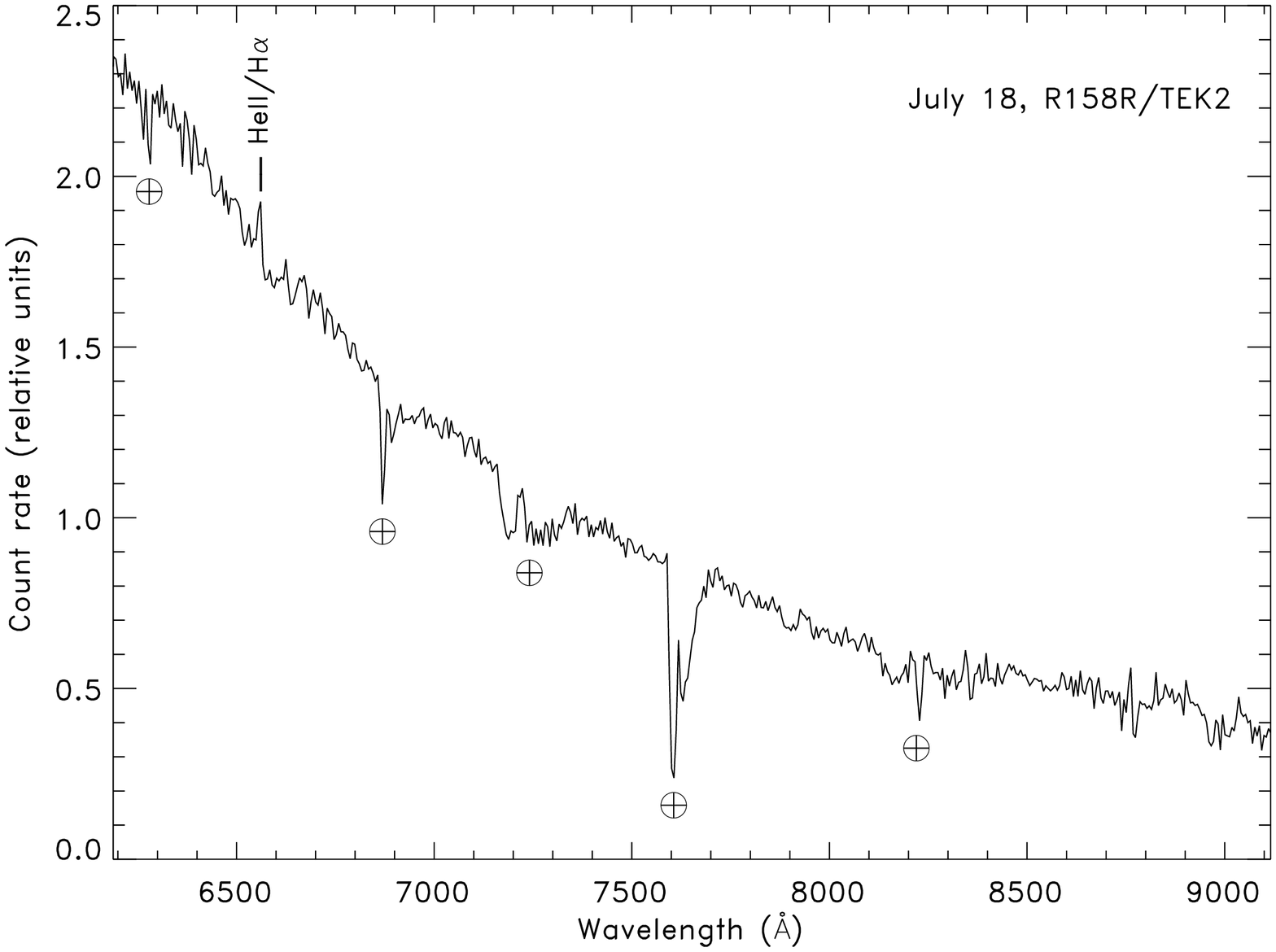}
\caption{Upper panels: average blue spectrum.  All identified
absorption or emission lines are marked.  Lower panel: red spectrum,
binned $\times2$ for clarity.}
\label{LineSpecFig}
\end{figure}

Standard {\sc iraf} procedures were used to de-bias the images.  Small
scale CCD sensitivity variations were removed using a high
signal-to-noise flat field obtained at the beginning of the night of
July 20.  Fringing (significant above $\sim5900$\,\AA) and
illumination variations were removed using contemporaneous flats taken
at the same time as arc calibrations.  One-dimensional spectra were
extracted from the processed images using the optimal extraction
method (Horne 1986; Marsh 1989).

Wavelength calibration was interpolated between contemporaneous
exposures of a copper-argon arc lamp.  Second-order flexure effects
were corrected using the O\,{\sc [I]} 5577.37\,\AA\ night
sky emission line.  This correction was always $\la 0.3$\,\AA.  We
performed flux calibration as a two-stage process.  The spectrograph
slit was aligned to pass through another star to serve as a
comparison.  Spectra of both stars were extracted from each image, and
the spectrum of XTE\,J2123--058 was then calibrated relative to the on-slit
comparison star in each case.  A final image using a 4\,arcsec slit
was used to calibrate the comparison star relative to the
spectrophotometric standard Feige 110 (Oke 1990).  The absolute flux
calibration is sensitive to uncertainty in the extinction curve.  The
wide slit observation used for absolute calibration was obtained at
airmass 1.72, whereas the spectrophotometric standard, Feige 110, was
observed at airmass 1.25.  An average extinction curve for La Palma
(King 1985) was assumed in correcting for this difference but the
presence of significant dust in the atmosphere at the time of the
observations means that the extinction curve may be far from
average and that the extinction may also depend on the azimuth of the star.
Differential slit losses represent an additional problem as they vary
between spectra.  This difficulty will be discussed in Sect.\
\ref{ContSect}.

To obtain the average blue spectrum from July 19--21 shown in the
upper two panels of Fig.\ \ref{LineSpecFig}, two combined spectra were
generated.  The first was a straight sum of recorded counts before
slit loss and extinction corrections.  As the slit losses in
particular are strongly dependent on wavelength and time, this
improves the signal-to-noise ratio dramatically compared to averaging
calibrated spectra.  The second was an average of calibrated spectra
resampled onto a uniform phase grid.  This should give a true average
energy distribution.  This second spectrum was then used to calibrate
the first high signal-to-noise one.  The main drawback of this process
is that relative fluxes of widely separated lines may be subject to
systematic errors.  For example, the line structure in the average
blue-end spectrum is predominantly determined from the phase range
$\sim$0.0--0.5 when slit losses were minimised, whereas the red-end
line spectrum is closer to a uniformly weighted average.  Our single
red spectrum from July 18/19 is also shown in the lower panel of Fig.\
\ref{LineSpecFig}.
%
%
\section{The continuum flux distribution and light curve}
\label{ContSect}
\begin{figure}
\epsfig{angle=90,width=3in,file=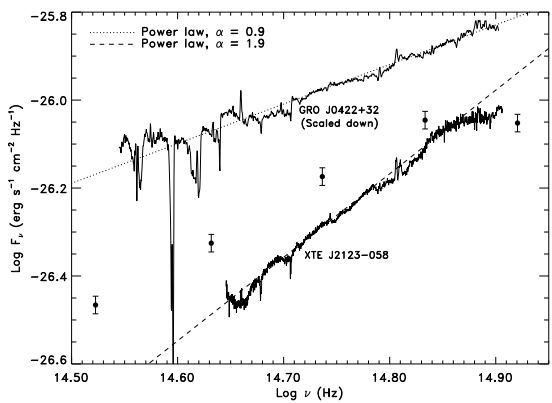}
\caption{Comparison of dereddened spectral energy distributions of
XTE\,J2123--058 and GRO\,J0422+32.  Points indicate the distribution
inferred earlier in the outburst (June 30) from photometry by Tomsick
et al.\ (1998a).  Approximately fitting power-laws are also shown.
The spectral energy distribution of XTE\,J2123--058 appears steeper
than that of GRO\,J0422+32.}
\label{ContSpecFig}
\end{figure}

We show the dereddened spectral energy distribution of XTE\,J2123--058
in Fig.\ \ref{ContSpecFig}.  For comparison, we also show the
photometry of Tomsick et al.\ (1998a) and a spectrum of the black hole
candidate GRO\,J0422+32 in outburst (Shrader et al.\ 1994).
Dereddening assumes $E(B-V)=0.12$ for XTE\,J2123--058 (Sect.\
\ref{ISLinesSection}) and $E(B-V)=0.3$ for GRO\,J0422+32 (as discussed
in Hynes \& Haswell 1999).  Power-laws of spectral index $\alpha=1.9$
for XTE\,J2123--058 and $\alpha=0.9$ for GRO\,J0422+32 (where $F_{\nu}
\propto \nu^{\alpha}$) are overplotted to characterise the spectral
energy distributions (SEDs) of the two objects.  A comparison of these
objects is appropriate as GRO\,J0422+32 has an orbital period of
5.1\,hr and is the closest black hole analogue of XTE\,J2123--058.
Its SED is representative of other short period black hole systems,
but is redder than that of XTE\,J2123--058 (compared to both
spectroscopy and photometry); the latter SED is closer to an
$\alpha=2$ Rayleigh-Jeans spectrum.  This suggests that the emission
regions in XTE\,J2123--058 are hotter than in GRO\,J0422+32, although
the possibility of errors in extinction correction, discussed in
Sect.\ \ref{ReductionSection} weaken this conclusion.

\begin{figure}
\epsfig{width=3in,file=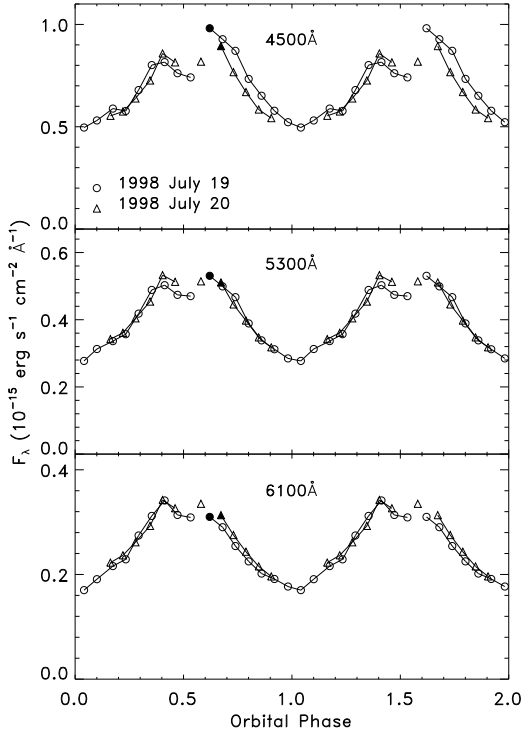}
\caption{Continuum light curves of XTE\,J2123--058 at several
wavelengths.  Joined points indicate continuous sequences from Table
\ref{WHTTable}.  Solid points indicate the first observation of each
night; each observing night ran approximately from phase 0.6--1.6.
Apparent inconsistencies between the two nights, and differences
between different wavelengths are likely due to calibration errors.}
\label{ContLCFig}
\end{figure}

We show in Fig.\ \ref{ContLCFig} light curves for three `continuum'
bins at 4500\,\AA, 5300\,\AA\ and 6100\,\AA.  The light curves show
very similar shapes, with no significant differences in profile or
amplitude within this wavelength range.  The apparent differences
between them, most noticeable around phase 0.6, are likely due to
calibration uncertainties: coverage on each night was approximately
from phase 0.6 through 1.6.

The light curve morphology is fully discussed in Paper I.  We believe
it is mainly due to the changing aspect of the heated companion star.
This is the dominant light source at maximum light, near phase 0.5.
At minimum light (phase 0.0) the heated face is obscured and we see
the accretion disc only.  At all phases the unilluminated parts of the
companion star are expected to contribute negligible flux as the
outburst amplitude is $\sim 5$ magnitudes.  We see no strong
dependence of this modulation amplitude on wavelength.  This is
expected if the emission regions are sufficiently hot that we only see
the $F_{\nu} \propto \nu^{2}$ Rayleigh-Jeans part of the spectrum and
is consistent with the very steep SED shown in Fig.\
\ref{ContSpecFig}.
%
%
\section{The line spectrum}
\label{LineSection}
At first glance, XTE\,J2123--058 presents a nearly featureless
spectrum, with only the Bowen blend (\NIII/\CIII\ 4640\,\AA) and
\HeII\ 4686\,\AA\ prominent.  A number of weaker emission lines are
present too, however, with complex Balmer line profiles and weak
interstellar absorption features also seen.

\begin{table}
\caption{Spectral lines detected in XTE\,J2123--058.  No equivalent
widths are calculated for Balmer lines due to the complex nature of
the line profiles, and for higher order \HeII\ lines due to the
difficulty of setting a meaningful continuum level.  Errors in
equivalent widths are statistical errors in the average spectrum and
do not account for intrinsic variability.}
\label{EmissionTable}
\begin{tabular}{lll}
\noalign{\smallskip}
\hline
\noalign{\smallskip}
Identification      & Wavelength & Equivalent  \\
                    & (\AA)      & width (\AA)\\
\noalign{\smallskip}
\CaII\,K (IS)          & 3933.7        &$+0.30\pm0.04$ \\ 
\HeII\ 12--4/H$\delta$ & 4100.0,4107.7 &               \\ 
\HeII\ 11--4           & 4199.8        &               \\ 
\HeII\ 10--4/H$\gamma$ & 4338.7,4340.5 &               \\ 
DIB (IS)               & 4428          &$+0.24\pm0.05$ \\
\HeII\ 9--4            & 4541.6        &               \\ 
\NIII/\CIII            & 4640          &$-1.98\pm0.03$ \\ 
\HeII\ 4--3            & 4685.7        &$-1.92\pm0.03$ \\ 
\HeII\ 8--4/H$\beta$   & 4859.3,4861.3 &               \\
\HeII\ 7--4            & 5411.5        &$-0.74\pm0.05$ \\ 
\CIV                   & 5801.5,5812.1 &$-0.39\pm0.06$ \\ 
\NaI\,D (IS)           & 5890.0,5895.9 &$+0.59\pm0.03$ \\ 
\HeII\ 6--4/H$\alpha$  & 6560.1,6562.5 &               \\ 
\noalign{\smallskip}
\hline
\end{tabular}
\end{table}

\subsection{\NIII-\CIII\ and \HeII\ 4686\,\AA\ emission}
The equivalent widths of the \NIII/\CIII\ Bowen blend and \HeII\
4686\,\AA\ imply a \NIII/\CIII-\HeII\ ratio of $1.03\pm0.02$,
thoroughly consistent with the ratios observed in other Galactic LMXBs
(typically $\sim0.5-1.5$; Motch \& Pakull 1989).  In particular, this
ratio is significantly higher than seen in LMXBs in low-metallicity
environments (e.g.\ $<0.12$ for LMC~X-2 and $<0.20$ for 4U\,2127+11 in
the low-metallicity globular cluster M15; Motch \& Pakull 1989).
This suggests that XTE\,J2123--058 originates in the Galactic
population; see Sect.\ \ref{OriginSection} for further discussion.  Note
that as these two lines are very close, this ratio is not affected by
the uncertainty in line flux calibration noted in Sect.\
\ref{ReductionSection}.
\subsection{\HI\ Balmer and \HeII\ Pickering lines}
The Balmer lines exhibit a complex profile with broad absorption and a
structured emission core.  This broad absorption plus core emission
line structure is common in SXTs in outburst.  Balmer line profiles in
GRO\,J0422+32 (Shrader et al.\ 1994; Callanan et al.\ 1995) look very
similar, showing a double-peaked core in broad absorption.

\begin{figure}
\epsfig{width=3in,file=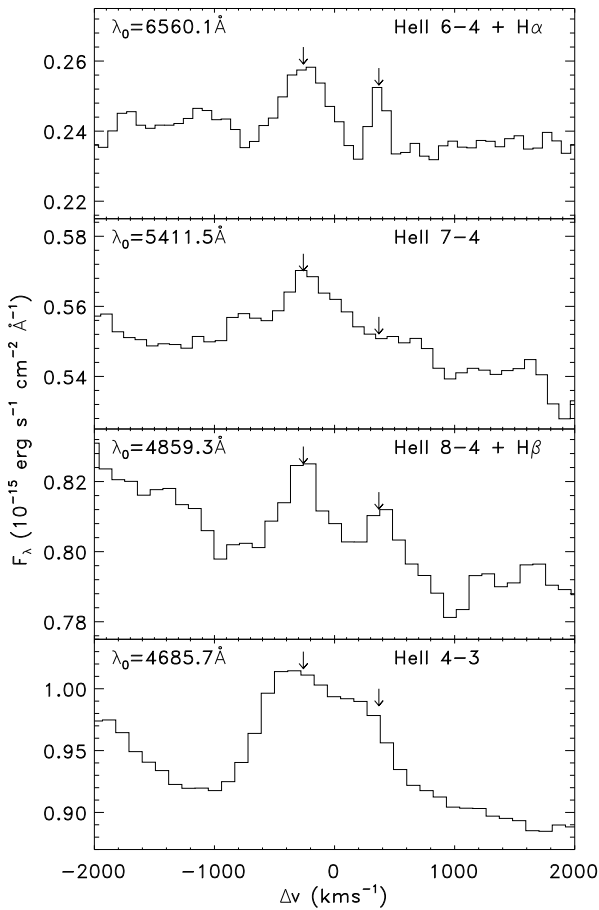}
\caption{Upper three panels: Selected \HeII\ Pickering lines and
coincident \HI\ Balmer lines.  The horizontal axis is velocity with
respect to the rest velocity of the \HeII\ line in each panel.  The
arrows mark the two prominent emission components and are drawn at the
same velocities in each plot.  The left-hand peak is clearly a \HeII\
feature, the right hand one is probably due to Balmer lines as it only
appears where Balmer lines are present.  The spectrum has been
dereddened assuming $E(B-V)=0.12$ and the vertical flux scale is the
same in each of the upper three plots.  Emission line fluxes can
therefore be directly compared.  The rest wavelength of the Balmer
lines corresponds to velocities of $+100$\,km\,s$^{-1}$ for \Halpha\
and $+130$\,km\,s$^{-1}$ for \Hbeta.  The lower panel shows \HeII\
4686\,\AA\ on the same velocity scale for comparison.  It is clearly
broader than the Pickering lines, especially on the red wing, but the
peak velocity is similar.}
\label{BalmerFig}
\end{figure}

The interpretation of the profiles is complicated by the presence of
coincident \HeII\ Pickering lines ($n=$ 5,6,7\ldots to 4 transitions)
within 2\,\AA\ of each Balmer line.  The cores are compared in Fig.\
\ref{BalmerFig}.  We show H$\alpha$ and H$\beta$, together with the
\HeII\ 6--4, 7--4 and 8--4 lines (Pickering $\beta$--$\delta$).  The
\HeII\ 4686\,\AA\ line is also shown for comparison.  Higher order
lines show similar behaviour, but with lower resolution and signal-to-noise.
The two peaks marked are clearly present in both H$\alpha$ and
H$\beta$, and can also be identified in H$\gamma$ and H$\delta$.  The
blue peak is also seen in \HeII\ 5411\,\AA\ (7--4), but the red peak
apparently is not.  Peak 4686\,\AA\ emission also coincides with the
blue peak.  The blue peak, however, appears weaker in flux in \HeII\
5411\,\AA\ than in H$\alpha$ or H$\beta$.  As noted in Sect.\
\ref{ReductionSection} we should be cautious about systematic errors
in comparing widely separated lines, and the difference is not large.
We therefore cannot rule out the possibility that the blue peak is
purely \HeII\ emission.  Our suggested decomposition of the line
profile is therefore that the blue peak contains \HeII\ emission, and
possibly also some Balmer emission.  The red peak is pure Balmer
emission.  The broad absorption also appears to be a Balmer feature.

There is a further complication.  As will be described in more detail
in Sect.\ \ref{LineVarSection}, H$\alpha$, and possibly also H$\beta$,
are also subject to transient absorption, strongest in the phase range
0.35--0.55.  The average \Halpha\ line profile in this phase range is
shown in Fig.\ \ref{AbsorptionFig}.  As the absorption component
clearly lies between the two peaks, it may be that weaker absorption
at all phases is partly responsible for the apparent separation of the
two peaks.
\subsection{Interstellar lines}
\label{ISLinesSection}
We measure equivalent widths for the Na\,D lines of $0.29\pm0.02$ and
$0.30\pm0.02$\,\AA.  Based on the strength of the D1 line and the
calibration of Munari \& Zwitter (1997) we estimate $E(B-V) = 0.12 \pm
0.05$.  Assuming $R_{\rm V}=3.1$, this implies $A_{\rm
V}=0.37\pm0.15$, agreeing with the value of $A_{\rm V}=0.36$ quoted by
Homan et al.\ based on dust maps in Schlegel, Finkbeiner \& Davis
(1998).  We note, however, that Soria et al.\ (1999) estimate a
somewhat lower value of $E(B-V)=0.054$ using the reddening maps
derived by Burstein \& Heiles (1982) from \HI\ column densities and
galaxy counts.
%
%
\section{Line variability}
\label{LineVarSection}
\begin{figure}
\epsfig{width=3in,file=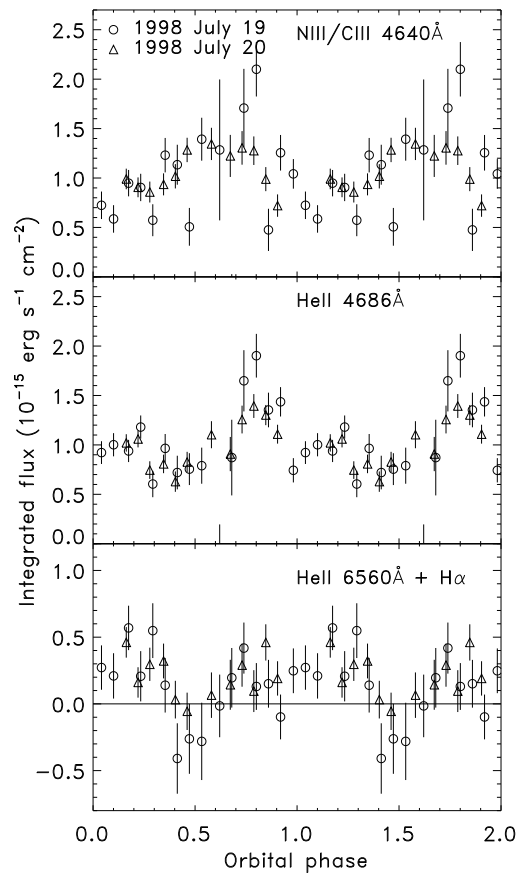}
\caption{Emission line light curves of XTE\,J2123--058.  Line fluxes
are measured relative to the continuum, so the absorption dip in the
\HeII\ 6560\,\AA/\Halpha\ light curve is real absorption below the
continuum, not simply obscuration of the emission component; this can
be seen more clearly in Fig.\ \ref{AbsorptionFig}}
\label{LineLCFig}
\end{figure}

\begin{figure*}
\epsfig{width=6in,file=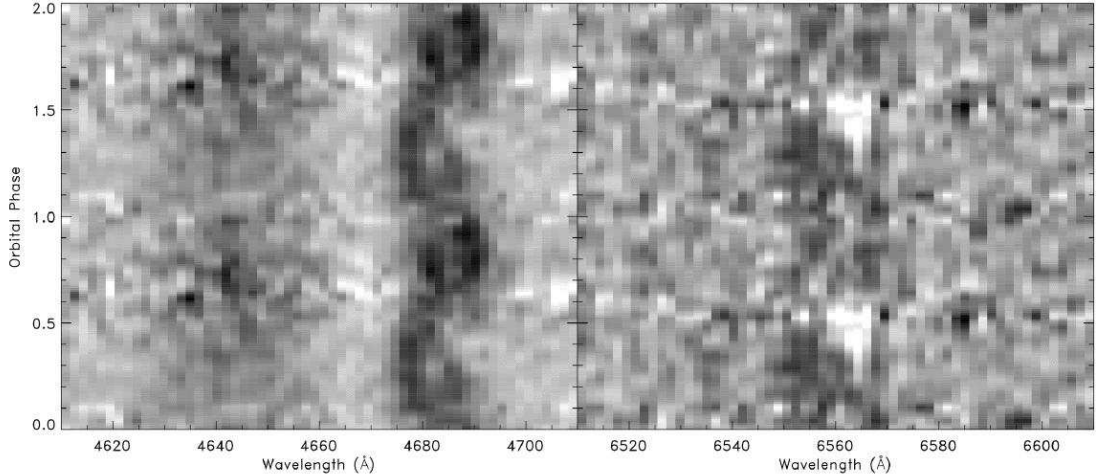}
\caption{Trailed spectrograms of XTE\,J2123--058.  Data have been
phase-binned, with the same data plotted twice.  The left hand panel
shows behaviour of \CIII/\NIII\ 4640\,\AA\ and \HeII\ 4686\,\AA\
emission.  The right hand panel shows blended \HeII\ 6560\,\AA\ and
\Halpha\ behaviour.  Note the phase 0.5 \Halpha\ absorption component;
the white spot in the right hand panel. }
\label{SpectrogramFig}
\end{figure*}

The emission lines show significant changes over an orbital cycle,
with differences in behaviour between lines.  \NIII/\CIII\ 4640\,\AA\
(Bowen blend) shows a flux modulation, as does \HeII\ 4686\,\AA\
emission which also shows complex changes in line position and
structure.  \HeII\ 6560\,\AA\ and \Halpha\ are blended and in addition
to probably showing flux and wavelength modulations are subject to a
transient absorption feature that is strongest around phase 0.5.  The
changes are illustrated with integrated light curves in Fig.\
\ref{LineLCFig} and with trailed spectrograms in Fig.\
\ref{SpectrogramFig}.

The Bowen blend light curve appears similar to the continuum light
curves shown in Fig.\ \ref{ContLCFig}.  Bowen emission may therefore
originate on the heated face of the companion star, with the modulation
arising from the varying visibility of the heated region.  

\HeII\ light curves show a double peaked structure with a strong peak
near phase 0.75 and a weaker one near 0.25.  We should be cautious in
interpreting this light curve in detail, however, as the complex line
behaviour seen in the trailed spectrogram indicates multiple emission
sites; the integrated light curve is an average of different light
curves of several regions.

\begin{figure}
\epsfig{angle=90,width=3in,file=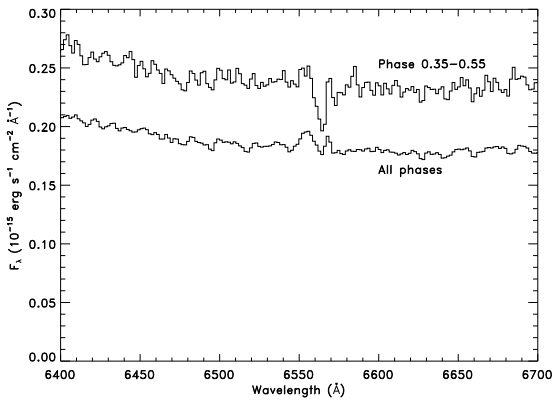}
\caption{Average spectra near \Halpha\ (lower) together with average
of phases 0.35--0.55 alone (upper).  Real absorption can clearly be
seen in the upper spectrum.}
\label{AbsorptionFig}
\end{figure}

\Halpha\ emission is blended with \HeII\ 6560\,\AA, and as discussed
in Sect.\ \ref{LineSection} also comprises several components.  The
most striking is the transient absorption component seen near phase
0.5.  This can be clearly seen in the light curve and the trailed
spectrogram.  The light curve is deceptive as it also reflects changes
in the emission flux; it suggests that the absorption feature is
centred at phase 0.5, but, as can be seen from the trailed
spectrogram, this is not really the case.  The feature begins quite
sharp and red, around phase 0.3 and then becomes progressively more
extended to the blue, fading out by phase 0.6.  Whilst it is sometimes
difficult to distinguish an absorption component from an absence of
emission, that is not a problem here; the absorption definitely
extends below the continuum as can clearly be seen in Fig.\
\ref{AbsorptionFig}.

%
%
\section{Doppler tomography}
\begin{figure}
\begin{center}
\epsfig{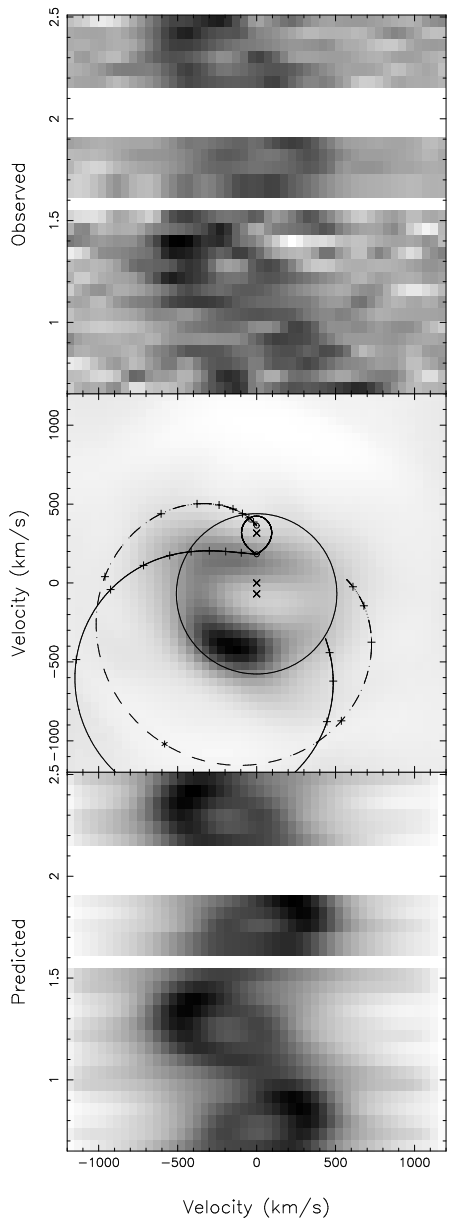}
\end{center}
\caption{Doppler tomogram of \HeII\ 4686\,\AA\ emission in
XTE\,J2123--058.  The upper panel shows the observed trailed spectra,
normalised to constant total line flux.  The middle panel shows the
tomogram itself, and the lower panel shows the ideal trailed
spectrogram reconstructed from the tomogram.  In the middle panel, the
solid arc is the ballistic stream trajectory, the dashed arc the
Keplerian (disc) velocity at the stream position and the large circle
is the Keplerian velocity at the outer disc edge.  Phases below 1.6
come from the first night, those above this from the second; three
complete cycles have been omitted for clarity.}
\label{TomogramFig}
\end{figure}

We have used the technique of maximum entropy Doppler tomography
(Marsh \& Horne 1988) to identify \HeII\ 4686\,\AA\ emission sites in
velocity space.  Our results are shown in Fig.\ \ref{TomogramFig}.
The analytical back-projection method gave similar results.  We should
beware, however, that one of the fundamental assumptions of Doppler
tomography, that we always see all of the line flux at {\it some}
velocity, is clearly violated, as the integrated line flux is not
constant.  We attempt to account for this by normalising the spectra
(dividing by a smooth fit to the \HeII\ 4686\,\AA\ light curve in
Fig.\ \ref{LineLCFig}).  If we do not normalise then the structure of
the derived tomogram is very similar, so our results do not appear to
be sensitive to this difficulty.

Ideally, to perform Doppler tomography we should know the systemic
velocity of the binary.  We have no {\it a priori} knowledge of this,
so we estimate it from the \HeII\ 4686\,\AA\ line itself.  This is
done by repeating the tomographic analysis for different assumed mean
velocities and seeking to minimise the residuals of the fit of the
reconstructed spectrogram to the data (Marsh \& Horne 1988).  Our
measured heliocentric systemic velocity is $\sim-90$\,km\,s$^{-1}$ The
implications of this velocity are discussed in Sect.\
\ref{OriginSection}.

In order to interpret the tomogram quantitatively, we also require
estimates of system parameters.  These were derived from results of
fits to outburst light curves (see Paper I).  The parameters we assume
are given in Table \ref{ParamTable}.  The masses deserve comment.  The
light curve fits only provide a significant constraint on the mass
ratio, of $q = M_1 / M_2 = 4.6 ^{+0.5} _{-0.2}$.  To obtain actual
masses, one must make physical assumptions.  We assume that the
neutron star has a typical mass of $M_1 \sim 1.4$\,M$_{\sun}$ (see
Thorsett \& Chakrabarty 1999 for a review of neutron star masses) and
hence that the companion should have a mass of 0.3\,M$_{\sun}$.  As
discussed in Paper I, a main sequence companion should have a mass of
0.6\,M$_{\sun}$ so it is likely that the companion is evolved and
undermassive.  Other SXTs do appear to have undermassive companions
(Smith \& Dhillon 1998), so this is a very plausible scenario, and one
which is consistent with theoretical predictions (King, Kolb \&
Burderi 1996) that short period neutron star LMXBs should {\em only}
be transient if they have evolved companions.

\begin{table}
\caption{Adopted system parameters.  These are taken from Paper I
except where discussed in the text.}
\label{ParamTable}
\begin{tabular}{ll}
\hline
\noalign{\smallskip}
$P_{\rm orb} $ & 0.24821\,d \\
$i$            & 73$^{\circ}$ \\  
$M_1$          & 1.4\,M$_{\sun}$ \\
$M_2$          & 0.3\,M$_{\sun}$ \\
$R_{\rm L_1}$  & $6.6\times10^{10}$\,cm \\
$R_{\rm disc}$ & 0.75 $R_{\rm L_1}$ \\
\noalign{\smallskip}
\hline
\end{tabular}
\end{table}

The dominant emission site (corresponding to the main S-wave) appears
on the opposite side of the neutron star from the companion.  For any
system parameters, it is inconsistent with the heated face of the
companion and the stream/disc impact point.  There is a fainter spot
near the $\rm L_1$ point, corresponding to the fainter S-wave, which
is possibly associated with the accretion stream.  There also appears
to be a loop connecting these two spots and a fainter arc on the right
completing the circle.  As noted above, however, the assumption that
all the line flux is always visible at some velocity is not true for
these data, and normalising the total flux is only an approximate
solution when several emission components are present.  It is likely,
therefore, that some low-level artefacts are present and so weaker
features in the tomogram may not represent real emission sites.  To
test this, we have also generated tomograms in which data along the
S-wave corresponding to the bright spot are given a very low weight
(see Steeghs et al.\ 2000).  This should prevent artefacts due to
misfitting of the time-dependence of this feature.  We find that the
upper left loop appears robustly, and is likely to be a real
structure.  The apparent arc completing the circle on the right does
not appear and so is probably spurious.  It is notable that nearly all
emission has a lower velocity than Keplerian material at the outer
disc edge could have, and so if associated with the disc requires
sub-Keplerian emission.
%
%
\section{Discussion}
\subsection{A neutron star SW Sex system?}
In several respects XTE\,J2123--058 resembles the SW~Sex class of
cataclysmic variables.  These are all novalike variables, typically
seen at a relatively high inclination.  Their main defining
characteristics are (Thorstensen et al.\ 1991; Horne 1999; Hellier
2000):
\begin{enumerate}
\item
Balmer lines are single-peaked and their motions lag behind that of
the white dwarf.  In a Doppler map emission is typically (though not
always) concentrated in the lower-left quadrant at velocities lower
than any part of the disc.  A high velocity component is detectable
with velocity semi-amplitude $\sim 1000$\,km\,s$^{-1}$.  This has the
same phasing as the low velocity component; and their separate
identity is vigorously debated.
\item
Transient absorption lines are seen.  These are strongest near phase
0.5 or slightly earlier and move from red to blue between phases 0.2
and 0.6.
\item
Continuum eclipses have a V-shaped profile implying a flatter
temperature distribution than expected theoretically.  Balmer line
eclipses are shallow and occur earlier than continuum eclipses.  Red
shifted emission remains visible at mid-eclipse.
\end{enumerate}

On the first two points XTE\,J2123--058 shows striking similarities to
the SW~Sex systems.  Anomalous emission line behaviour is most clearly
seen in \HeII\ 4686\,\AA\ in this higher excitation system.  This line
appears, on average, single peaked.  The emission is concentrated in
the lower-left quadrant at sub-Keplerian velocities.  A high velocity
component is not detected, but cannot be ruled out without higher
quality data.  Transient absorption is seen in the Balmer lines before
and around phase 0.5.  It begins slightly redshifted and moves to the
blue.  XTE\,J2123--058 is the SXT with the highest known inclination
($i=73^{\circ}\pm4^{\circ}$; Paper I), strengthening the similarity to
SW~Sex systems.  It does not deeply eclipse, however, so it is
impossible to comment on the third point.

Various models have been proposed for the SW~Sex phenomenon including
bipolar winds, magnetic accretion and variations on a stream overflow
theme.  Current opinion favours the latter interpretation, with the
stream either being accelerated out of the Roche lobe by a magnetic
propeller, or re-impacting the disc.  Both of these models are
discussed in more detail below.
\subsection{Is this behaviour typical of LMXBs?}
The only other SXT for which outburst Doppler tomography of adequate
quality is available is the black hole candidate GRO\,J0422+32 (Casares
et al.\ 1995).  Using ephemerides determined subsequently (Harlaftis
et al.\ 1999), emission in this object appears to originate from the
accretion stream/disc impact point.  These data were obtained in the
December 1993 mini-outburst, $\sim500$\,d after the peak of the
primary outburst.

Amongst other short-period neutron star LMXBs in an active state, only
2A\,1822--371 (Harlaftis, Charles \& Horne 1997) has Doppler
tomography.  This was observed in \Halpha\ and exhibited disc emission
enhanced towards the stream impact point.  Similar behaviour is
suggested by a radial velocity analysis of \HeII\ (Cowley, Crampton \&
Hutchings 1982).  This is different to what we see in XTE\,J2123--058.
Radial velocity analyses of other systems, however, suggest similar
behaviour to that seen here.  4U\,2129+47 shows a very similar blue
spectrum to XTE\,J2123--058 (Thorstensen \& Charles 1982).  Clear
variations in the \HeII\ 4686\,\AA\ radial velocity are seen with
maximum radial velocity around phase 0.7--0.8 and semi-amplitude
$216\pm70$\,km\,s$^{-1}$.  Spectroscopy of EXO\,0748--676 (Crampton et
al.\ 1986) also reveals radial velocity modulations in the \HeII\
4686\,\AA\ line.  The radial velocity curve measured from the line
base shows maximum velocity at phase $0.76\pm0.16$ with semi-amplitude
$210\pm92$\,km\,s$^{-1}$.  For comparison, a similar radial velocity
curve fit to our \HeII\ spectra (Hynes et al.\ 1998) yielded maximum
velocity close to phase 0.75 with semi-amplitude
$\sim180$\,km\,s$^{-1}$.  Augusteijn et al.\ (1998) studied
4U\,1636--536 and 4U\,1735--44 and found the two systems similar to
each other.  Both show \HeII\ modulations with semi-amplitude
140--190\,km\,s$^{-1}$ with maximum radial velocities measured at
phases 0.87--1.04.\footnote{These phases have shifted by 0.5 with
respect to those quoted by the authors as they define phase 0 as
photometric {\em maximum}.}  We therefore suggest that the behaviour
we see is common, but not universal, in short period neutron star
LMXBs in a high state.
\subsection{The magnetic propeller interpretation}
\label{PropellerSect}
It is possible to explain the behaviour of the \HeII\ 4686\,\AA\
line in terms of the magnetic propeller model.  This model also
provides a possible mechanism for explaining the transient Balmer
absorption, although this is less satisfactory.

The essence of this model is that in the presence of a rapidly
rotating magnetic field, accreting diamagnetic material may be
accelerated tangentially, leading to it being ejected from the system.
It is easy to see how such a scenario may arise due to a rapidly
spinning, magnetised compact object.  This mechanism has been invoked
to explain the unusual behaviour of the cataclysmic variable AE~Aqr in
terms of a magnetic field anchored to a spinning white dwarf (Wynn,
King \& Horne 1997; Eracleous \& Horne 1996).  It is also likely that
magnetic propellers driven by neutron star magnetic fields exist, as
proposed by Illarianov \& Sunyaev (1975).  These may be very important
in suppressing accretion in neutron star SXTs in quiescence; see Menou
et al.\ (1999) and references therein for a discussion of this
situation.  There is an important difference of scale, however; spin
periods of white dwarfs are $\ga30$\,s; spin periods of neutron stars
are typically of order milliseconds.  For XTE\,J2123--058, $P_{\rm
spin} = 3.92\pm0.22$\,ms was suggested by Tomsick et al.\ (1999) based
on the beat-frequency model for kilohertz QPOs.  This identifies the
separation of the QPO peaks with the spin period.  While the peak
separation is not always exactly constant, Psaltis et al.\ (1999)
examine the model and conclude that the separation is still likely to
be close to the spin period.  Thus the spin period of the neutron star
in XTE\,J2123--058 is probably not far from 4\,ms.  For such rapid
rotation we must consider the light-cylinder, defined by the radius at
which the magnetic field must rotate at the speed of light to remain
synchronised with the compact object spin.  Outside of this radius,
the magnetic field will be unable to keep up and hence becomes wound
up.  This will make the propeller less effective, although it may
still produce some acceleration (Wynn et al.\ 1997; Wynn priv.\
comm.).  For a $\sim 4$\,ms spin period in XTE\,J2123--058, this is
$\sim6\times10^{-5}\,R_{\rm disc}$; for comparison, the closest
approach of a ballistic stream is $\sim0.2\,R_{\rm disc}$.  We would
therefore not expect a {\em neutron star} propeller to strongly affect
the stream overflow material, although simulations are needed to test
this.  An alternative mechansim was suggested by Horne (1999) for
SW~Sex systems, in which a magnetic propeller can arise from the
magnetic fields of the disc itself.  This has problems, in that it
should very efficiently remove angular momentum from the disc (Wynn \&
King priv.\ comm.), but this could perhaps be overcome if only a small
fraction of stream material passes through the propeller.  This
mechanism allows much lower angular velocities for the field and may
be the only {\em large scale} propeller mechanism possible in a
neutron star system.

We adopt the parameterisation used by Wynn, King \& Horne (1997) to
describe the magnetic propeller in AE~Aqr.  The magnetic field is
simplistically modelled as a dipole with angular velocity $\omega_{\rm
f}$.  While such a field may be a reasonable representation of that of
a compact star (white dwarf or neutron star), it is clearly a gross
simplification of that of a disc-anchored propeller.  In the latter
case the net effect would probably arise from a mixture of unstable
field structures of varying strengths and anchored at a range of
radii.  For simplicity and consistency with existing work on magnetic
propellers, however, we approximate the complex field structure with a
single rotating dipole.  The propeller mechanism assumes that the flow
breaks up into diamagnetic blobs of varying size and density.  The
field can only partially penetrate the blobs, and hence they are
accelerated, but not locked to the field lines.  The magnetic
acceleration on a blob is given by $g_{\rm mag} = -k[v-v_{\rm
f}]_{\perp}$, where $v$ is the stream velocity, $v_{\rm
f}=r\omega_{\rm f}$ is the field velocity and the component of the
velocity difference perpendicular to the field is taken.  The
parameter $k$ parameterises our ignorance of the magnetic field
strength and blob size and density, and following Wynn \& King (1995)
we adopt the functional dependence $k = k_0(r/r_0)^{-2}$, where $r_0 =
10^{10}$\,cm.  Different blob sizes and densities will lead to a
spread of values of $k_0$ and hence different trajectories.  The
uncertainty in the angular velocity of the magnetic field is also
factored into the uncertainty in $k_0$ provided it is high enough that
$v_{\rm f} \gg v$ at all points along the stream trajectory, and low
enough that all regions of interest are within the light cylinder.
For XTE\,J2123--058, this corresponds to a field period of
approximately $500\,{\rm s} \gg P_{\rm spin} \gg 10$\,s (i.e.\ a field
anchored at $0.16r_{\rm disc} \gg R \gg 0.012r_{\rm disc}$).  Within
this regime varying $P_{\rm spin}$ has the same effect as varying
$k_0$, and different trajectories can be described by a single
parameter.  This gives a plausible range of radii for anchoring the
magnetic field and so we adopt this simple case for our model and
arbitarily set $P_{\rm spin} = 50$\,s, corresponding to a field
anchored at radius $\sim0.03$ of the disk radius.  As noted above,
this is only an approximation to a more chaotic field structure.

As we are considering a system in which most of the accretion likely
proceeds through the disc, we assume that the propeller mechanism acts
only on overspill from the stream-disc impact point.  Our model
therefore calculates a trajectory which is purely ballistic until it
reaches the disc edge and is subject to a propeller force thereafter.
Quantitatively, this makes little difference to the results compared
to allowing the magnetic force to act everywhere on the stream.

\begin{figure*}
\begin{minipage}{3.5in}
\epsfig{height=3.5in,file=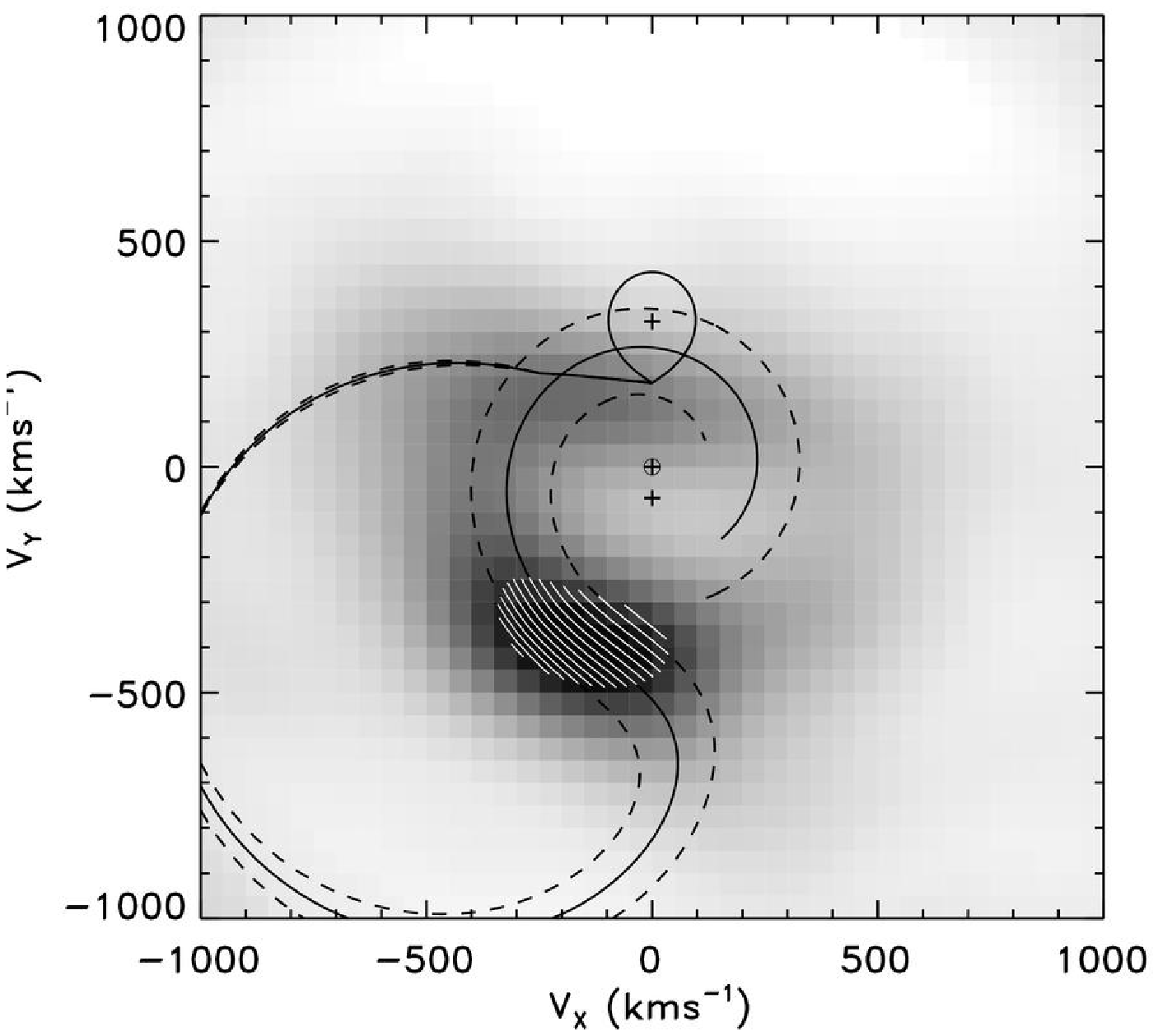}
\end{minipage}
\begin{minipage}{3.0in}
\vspace*{4mm}
\epsfig{height=3.0in,file=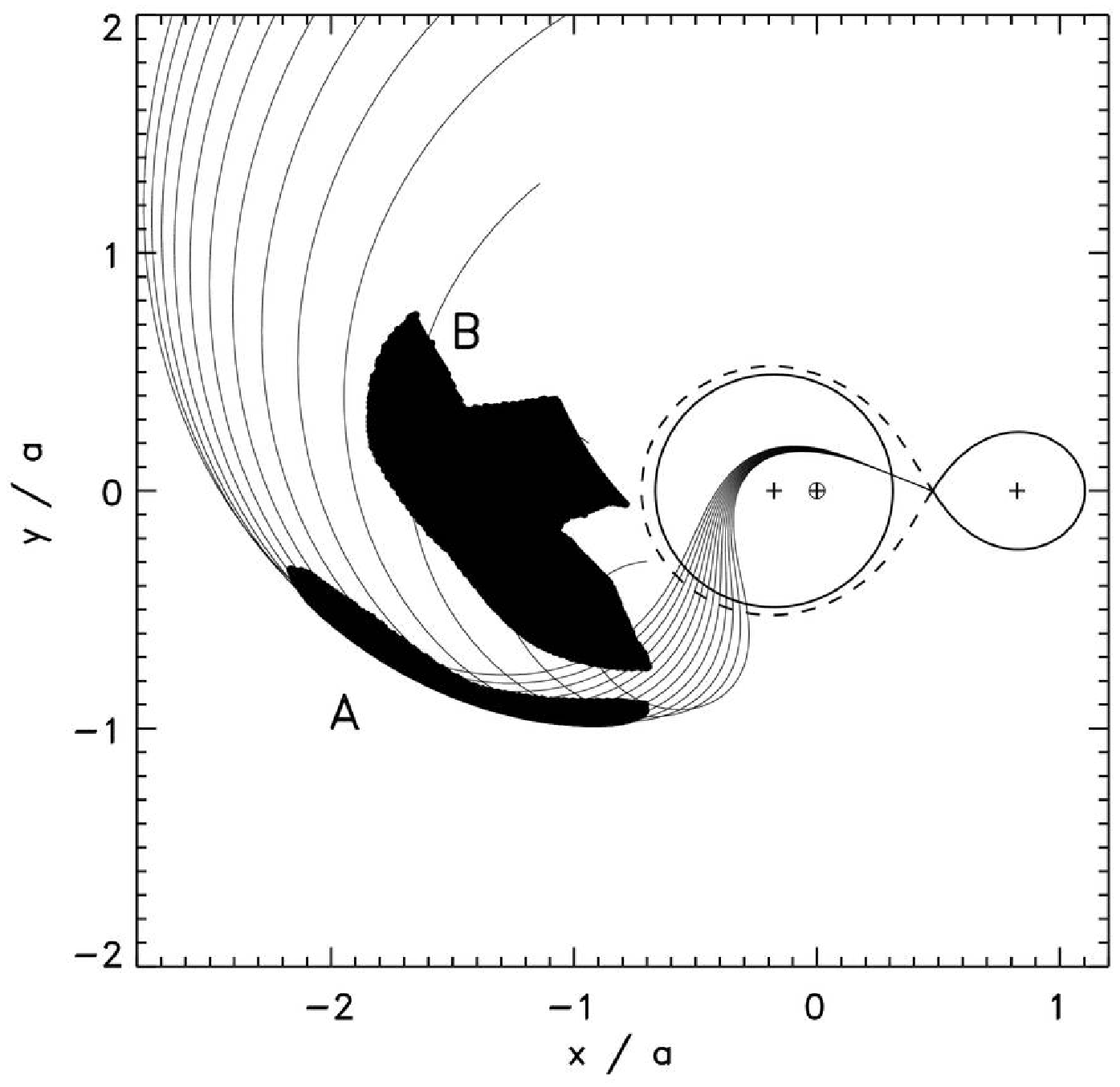}
\end{minipage}
\caption{A magnetic propeller model for \HeII\ emission; see text for
details of the model.  The left hand panel shows the tomogram with
three model trajectories corresponding to different values of $k_0$.
The white shading indicates the region of strongest emission.  In the
right hand panel, the trajectories are shown in real space in the
corotating frame; units are the binary separation.  Phase 0.0
corresponds to viewing from the right of the diagram, with 0.25 from
the bottom, 0.5 from the left and 0.75 from the top.  The region of
strongest emission in the tomogram maps to the solid area A, assuming
the propeller velocity field.  It is striking that this emission is
coincident with the location where the stream trajectories intersect
in real space.  Points within solid area B will obscure the disc with
the phasing and velocity observed in \Halpha.}
\label{PropellerFig}
\end{figure*}

We adjust the $k_0$ parameter to find a trajectory which passes
through the central emission on the tomogram.  This is shown in Fig.\
\ref{PropellerFig} together with two bracketting trajectories to show
the dependence on $k_0$.  For this to be a plausible model, we also
require an explanation for why one particular place on the trajectory
is bright.  Such an explanation is offered by Horne (1999): there is a
point outside the binary at which trajectories for different values of
$k$ intersect.  At this point, faster moving blobs cross the path of
slower blobs and enhanced emission might be expected.  This can be
seen in Fig.\ \ref{PropellerFig}.  We identify the region where the
trajectories collide with the emission region.

To make a quantitative comparison of our data with the model, we
transform the brightest emission spot of the tomogram into real space.
Such a conversion is not in general possible, of course, but can be
performed if system parameters and a velocity field are assumed,
provided that the velocity--space mapping is one--one or many--one.
That is true in this case if we only consider emission from the first
loop around the binary.  We construct a series of trajectories in
velocity space defined by uniform steps in $k_0$, and select points
for which the tomogram intensity is above a certain value.  The
selected points are shown spatially in the right hand panel of Fig.\
\ref{PropellerFig} as region A.  We can see that, subject to the
assumptions made, emission does occur at the point where trajectories
cross.  Our data are therefore consistent with the emission mechanism
suggested by Horne (1999).  We emphasise that this spatial map is only
valid in the context of the propeller model.

One point should be noted.  Hellier (2000) suggests that the propeller
model should predict a continuous S-wave, which is not seen.  This is
not necessarily the case.  If the region in which accelerated blobs
collide is optically thick in the line then emission is expected
predominantly from the inner edge of this region where fast moving
blobs impact slow moving blobs.  For the geometry we have considered,
as can be seen in Fig.\ \ref{PropellerFig}, this will result in
emission predominantly towards the top of the figure.  This
corresponds to seeing maximum emission near phase 0.75, exactly as is
observed in our data and in that of SW Sex systems, such as PX And
(Hellier \& Robinson 1994).  The phase dependence of emission line
flux is therefore not a problem with the propeller model, but instead
a prediction made by it.

We have demonstrated that both our \HeII\ tomogram and the line-flux
light curve are consistent with emission from the region where
material ejected by a propeller is expected to converge.  Can we
interpret the phase-dependent Balmer absorption as arising from the
same region, as was suggested by Horne (1999) in the context of SW~Sex
systems?  This model certainly predicts some ejecta in front of the
disc and/or companion at phases 0.3--0.6 where absorption is observed.
As can be seen in Fig.\ \ref{SpectrogramFig}, however, we also have
information on the absorption kinematics.  We can thus perform a
similar mapping to that used for the emission to determine if
absorption is predicted with the right phasing and kinematics and
where it would have to originate.  We construct a fine grid of
trajectories and for each point calculate the phase and velocity at
which this would pass in front of the centre of the disc.  If this
lies within the absorption region in the trailed spectrogram then we
mark that point as within region B in Fig.\ \ref{PropellerFig}.  We
find that the resulting points do populate the observed absorption
region fully, so the model can produce the absorption.  The material
involved, however, is not on the same trajectories as the emission,
but occurs on lower velocity tracks.  The absence of emission from
region B may indicate that only high velocity material undergoes
sufficiently energetic collisions to produce \HeII\ emission.  The
converse, that absorption does not reach to region A, could arise
because the emission region has a higher brightness temperature in the
lines than the disc continuum.  It also remains unexplained why
absorption is only seen along part of the trajectories.  For the low
velocity trajectories, some material may fall back onto the disc.  The
phase dependence may also be related to the height of the absorbing
material above the disc plane, as may the lack of absorption from
region A.  Clearly these are further questions to address in
developing a full propeller model for these data.  Such a
three-dimensional treatment is beyond the scope of our simplistic
model however, and we leave these points as observational requirements
that a full model must satisfy.
\subsection{The stream overflow and disc re-impact interpretation}
The most popular alternative explanation for the SW Sex phenomenon is
the accretion stream overflow and re-impact model.  This was
originally suggested by Shafter, Hessman \& Zhang (1988); for a recent
exposition see Hellier (2000).

In this model the overflowing stream can be thought of in two regions.
The initial part of the stream produces the transient absorption when
seen against the brighter background of the disc.  The latter part,
where the overflow re-impacts the disc, is seen in emission giving
rise to the high velocity component.  To reproduce observations, this
model requires a number of refinements.

In the simplest form of the model, the overflow stream should produce
absorption at all phases; it always obscures some of the disc.  This
problem is overcome by invoking a strongly flared disc so that the
overflow stream is only visible near phase 0.5.  We see an
absorption depth of 15\,per cent of the continuum flux.  At this
phase, at least 40\, per cent of the continuum flux actually comes
from the companion star, so we require absorption of 25\,per cent of
the visible disc light to achieve the observed depth, and hence at
least 25\,per cent of the visible disc to be covered by the obscuring
material; more if absorption is not total.  With a strongly flared
disc, however, the near side may be self-occulted so only
$\sim10$\,per cent of the total disc area need be covered by the
overflowing material.  This is not implausible, and simulations can
produce such a messy splash (Armitage \& Livio 1998).  There are
difficulties in invoking a strongly flared disc in XTE\,J2123--058,
however.  It has lower inclination than typical SW~Sex systems:
$73^{\circ}\pm4^{\circ}$ (Paper I) and does not show X-ray eclipses.
The strong photometric modulation due to heating of the companion also
constrains the amount that the disc can be flared; too much and the
companion will be shielded from X-rays.  In Paper I an upper limit to
the flare angle of $8\fdg6$ was derived (99\,per cent confidence) from
fitting of optical lightcurves $\sim10$\,days after the WHT
observations.  Comparing this with the viewing angle of
$17^{\circ}\pm4^{\circ}$ it is clear that the near side of the disc,
and hence the overflowing material, should never be completely
obscured, rendering the phase dependence difficult to account for.

The re-impact model requires that the strong emission in the
lower-left quadrant be produced by overlap of weaker emission from the
high-velocity component (the re-impact point) and the disc.  To
produce emission at the low velocities observed, however, requires the
`disc' component to be sub-Keplerian by a significant amount.  This
problem is not peculiar to XTE\,J2123--058, for which parameters are
uncertain, but common to the SW~Sex class in general.  In our case,
such a sub-Keplerian ring {\em may} be present, but it is likely that
this is not a real feature.  There is no evidence for a high velocity
component, but given our data quality, this is also inconclusive.  
\subsection{X-ray signatures of the absorbing material}
\label{XRaySection}
In principal X-ray observations can provide a diagnostic tool not
available in SW~Sex systems, and could constrain the location of the
absorbing material.  In the overflow and reimpact model it is
necessary for the disc to be highly flared to reproduce the phase
dependence of the absorption, and for the overflow to splash over a
moderately large area to reproduce the absorption depth.  One might
then expect that some of the overflowing material would cross our line
of sight to the neutron star and produce X-ray dips in the phase range
$\sim0.6$--1.0, as seen in other high inclination LMXBs, the dipping
sources.  Such dips are also possible in the propeller model, although
not necessary as this does not require a flared disc or extended
splash.  This model would, however, be expected to produce X-ray dips
in the same range as the transient \Halpha\ absorption, $\sim0.3$--0.6.

Tomsick et al.\ (1998b) reported no evidence for an X-ray orbital
modulation in {\it RXTE}/PCA data.  We have obtained archival PCA data
from 1998 July 14 and 22 and examined Standard-2 mode lightcurves to
reexamine this issue.  Background subtraction was done using bright
source background models dated 1999 September 17 and lightcurves were
extracted for both the whole PCA bandpass and for 3--5\,keV only.  We
find no evidence for dips of longer than 16\,s with depth greater than
10\,per cent at any orbital phase.  We have also phase-binned all of
these data to search for broader structures.  Approximately 60\,per
cent of phase bins have data combined from at least two binary orbits;
amongst these, with the exception of the X-ray bursts (Takeshima \&
Strohmayer 1998; Tomsick et al.\ 1998b) there are no excursions of
greater than $\pm10$\,per cent about the mean.
 
It is unfortunate that the {\it RXTE} data is not simultaneous with
our WHT observations and phase coverage is incomplete.  Thus we can
say that X-ray observations provide no evidence in favour of either
the reimpact or propeller models, but they do not rule either out.
\subsection{The origin of XTE\,J2123--058}
\label{OriginSection}
The question of the origin of XTE\,J2123--058 was raised in Sect.\
\ref{IntroSection}: is it intrinsically a halo object or was it kicked
out of the Galactic plane?  The strongest evidence is provided by the
ratio of \NIII/\CIII-\HeII.  This is typical of Galactic LMXBs and is
inconsistent with known halo objects.  In principal, the systemic
velocity is also an important clue.  For this object to have reached
its present location it must have received a kick when the neutron
star was formed, and hence it may now have a large peculiar
velocity (150--400\,km\,s$^{-1}$, Homan et al.\ 1999).  Our measured
{\em radial component} of velocity is $\sim-90$\,km\,s$^{-1}$ which is
not inconsistent with the predictions of Homan et al., which could
also be lowered if the source distance is less than the 10\,kpc value
assumed by those authors.  We should add a note of caution, however,
as systemic velocities inferred from emission lines are sometimes
unreliable; a definitive measurement of the systemic velocity must
await a study based on photospheric absorption lines of the companion
star.
%
%
\section{Conclusions}
Three key issues were raised in Sect.\ \ref{IntroSection} and we
conclude by returning to these.  As a neutron star transient,
XTE\,J2123--058 is valuable for comparison with black hole systems.
Of the latter, GRO\,J0422+32 is the most similar, and appears to show
a less steep optical spectral energy distribution than
XTE\,J2123--058, consistent with the suggestion of King et al.\ (1997)
that irradiation is more efficient in neutron star systems.  The high
orbital inclination of XTE\,J2123--058 has made possible Doppler
tomography of \HeII\ emission, and the detection of transient Balmer
line absorption.  These effects suggest similarities to SW Sex type
cataclysmic variables.  Unlike SW Sex systems, XTE\,J2123--058
shows characteristic SW Sex emission line kinematic effects (in \HeII)
decoupled from transient absorption providing constraints not possible
when both effects occur in the same line.  We have examined our
results in the context of stream overflow models applied to these
systems, both with stream reimpact and propeller ejection.  The
reimpact model, while plausible physically, has difficulties
accounting for our observations.  The propeller model can successfully
explain both emission and absorption behaviour, but has yet to
overcome theoretical difficulties.  Both theoretical work, and higher
quality observations will be required to conclusively decide between
the models, and it remains possible that neither is correct.  As an
object currently located in the Galactic halo, the origin of
XTE\,J2123--058 is also of interest.  Bowen line strengths, relative
to \HeII, indicate abundances similar to Galactic sources but
different to globular cluster and Magellanic LMXBs and hence we
suggest that this object was formed in the plane and subsequently
kicked out.
%
%
\section*{Acknowledgements}
RIH was initially supported by a PPARC Research Studentship and now by
grant F/00-180/A from the Leverhulme Trust.  Thanks to Chris Shrader for
providing the spectrum of GRO\,J0422+32 for comparison.  Doppler
tomography used {\sc molly}, {\sc doppler} and {\sc trailer} software
by Tom Marsh.  RIH would like to thank Danny Steeghs, and Daniel Rolfe
for enlightening discussions and useful software; Graham Wynn, Jan van
Paradijs, Keith Horne and Tom Marsh have also provided valuable
suggestions and advice, as has our referee, Coel Hellier.  The William
Herschel Telescope is operated on the island of La Palma by the Isaac
Newton Group in the Spanish Observatorio del Roque de los Muchachos of
the Instituto de Astrof\'\i{}sica de Canarias.  This research has made
use of the SIMBAD database, operated at CDS, Strasbourg, France, the
NASA Astrophysics Data System Abstract Service and of data obtained
through the High Energy Astrophysics Science Archive Research Center
Online Service, provided by the NASA/Goddard Space Flight Center.
%
%

%
\end{document}